\documentclass[twocolumn,prl,showpacs,preprintnumbers,amsmath,amssymb]{revtex4}

\usepackage{graphics}
\usepackage{color}
\usepackage{epsfig}

\usepackage{graphicx}
\usepackage{dcolumn}
\usepackage{bm}
\usepackage{multirow}

\begin{document}


\title{Universal Dimer in a Collisionally Opaque Medium:\\
Experimental Observables and Efimov Resonances}

\author{Olga Machtey, David A. Kessler}
\author{Lev Khaykovich}%
 \email{lev.khaykovich@biu.ac.il}
\affiliation{%
Department of Physics, Bar-Ilan University, Ramat-Gan, 52900 Israel
}%


\begin{abstract}
A universal dimer is subject to secondary collisions with atoms when
formed in a cloud of ultracold atoms via three-body recombination.
We show that in a collisionally opaque medium, the value of the
scattering length that results in the maximum number of secondary
collisions may not correspond to the Efimov resonance at the
atom-dimer threshold and thus can not be automatically associated
with it. This result explains a number of controversies in recent
experimental results on universal three-body states and supports the
emerging evidence for the significant finite range corrections to
the first excited Efimov energy level.
\end{abstract}

\maketitle Few-body systems with a resonantly enhanced two-body
scattering length $a$ display universal properties in the sense that
they are independent of the details of the short-range interaction
potential~\cite{Braaten&Hammer06}. In the two-body domain with
repulsive interactions, the most weakly bound energy level of the
universal dimer depends solely on $a$, scaling as $1/a^2$. The
central paradigm in the three-body domain, predicted in the early
$70$s by V.~Efimov~\cite{Efimov70}, is associated with the infinite
ladder of universal bound states with discrete scaling invariance.
Beside $a$, universal trimers depend on the three-body parameter,
which accounts for all the short-range physics not already included
in $a$. In recent experiments with ultracold atoms many aspects of
universality were verified mainly by localizing Efimov resonances
which are associated with the crossing points of the trimers'
binding energy levels with either the three-atom (for $a<0$) or the
atom-dimer (for $a>0$) continua (for a recent review see
Ref.~\cite{Ferlaino11}). Surprisingly, positions of the lowest
Efimov resonances for $a<0$ were found to be universally related to
the van der Waals lengths of the two-body interaction potentials in
different atomic
species~\cite{Kraemer06,Berninger11,Zaccanti09,Gross09,Gross10,Pollack09,Ottenstein08,Huckans09,Williams09,Lompe10,Wild11}
indicating a universal three-body parameter. This remarkable
experimental discovery has attracted intense theoretical attention
which suggests a whole new understanding of Efimov physics in
ultracold atoms~\cite{Wang11,Chin11}. However, the situation with
the Efimov resonances at the atom-dimer threshold (for $a>0$)
remains unclear. It appears to significantly deviate from the
universal picture in all species investigated to
date~\cite{Knoop09,Knoop09a,Lompe10,Nakajima10,Pollack09,Zaccanti09,Nakajima11}.
Suggestions have been made that the finite range of interaction
potentials might be responsible for these discrepancies. However,
here we show that in some cases a different effect, directly related
to the specifics of the experimental observables, can cause a
significant shift in the resonances' positions.

To show this effect we start by elaborating on two
experimental strategies developed recently to localize an Efimov
resonance at the atom-dimer threshold, denoted in the following as
$a_{*}$. Both are based on the theoretical prediction that resonant
enhancements of the atom-dimer elastic ($\sigma_{e}$) and inelastic
($\sigma_{i}$) cross sections are expected in the vicinity of
$a_{*}$, according to the following analytical
expressions~\cite{Braaten&Hammer06,Braaten&Hammer07}:
\begin{eqnarray}
\sigma_{e}&=&84.9\frac{\sin^{2}\left[s_{0}\ln(a/a_{*})+0.97\right]+\sinh^{2}\eta_{*}}
{\sin^{2}\left[s_{0}\ln(a/a_{*})\right]+\sinh^{2}\eta_{*}}a^{2} \label{eq:sigmae}\\
\sigma_{i}&=&\frac{20.3}{3v}\frac{\sinh (2\eta_{*})}{\sin^{2}
\left[s_{0}\ln(a/a_{*})\right]+\sinh^{2}\eta_{*}}\frac{\hbar
a}{m_{A}}\;, \label{eq:sigmai}
\end{eqnarray}
where $s_{0}=1.00624$, $\eta_{*}$ is the lifetime of the Efimov
state, $v$ is the velocity of the dimer and $m_{A}$ is the mass of
the atom~\footnote{We use the correct sign before $0.97$ in
Eq.~(\ref{eq:sigmae}) as it appears in Ref.~\cite{Braaten&Hammer07}
as opposed to Ref.~\cite{Braaten&Hammer06}.}. In the first strategy,
realized with ultracold Cs and $^{6}$Li, an atom-dimer mixture is
initially prepared in a shallow optical trap and the decay rate of
dimers is then monitored as a function of
$a$~\cite{Knoop09,Knoop09a, Lompe10,Nakajima10}. Thus, in this
approach, the maximum of the enhancement in the atom-dimer
\emph{inelastic} collision rate is expected to coincide with
$a_{*}$. In the second strategy, pioneered with
$^{39}$K~\cite{Zaccanti09} and then used with
$^7$Li~\cite{Pollack09,Machtey11}, the three-body recombination
induced loss rate of atoms is measured. In each recombination event
a universal dimer is formed with initial kinetic energy equal to one
third of its binding energy ($E_{d}/3$), which is usually much
higher than the shallow optical trap depth. Moreover, the dimer is
formed with the largest probability at the trap center where the
density is highest. Then, on its way out, it may undergo secondary
collisions with atoms, eject them out of the trap and cause the
number of lost atoms per recombination event to greatly exceed
three. Therefore, in this approach, the position of the maximum of
the atom-dimer \emph{elastic} collision rate is anticipated to
reveal $a_{*}$.

In this Letter we consider the problem of secondary collisions of
dimers with atoms and develop a model to calculate their mean number as
a function of $a$. Our central message is that the elastic and the
inelastic processes are intimately interrelated and can not be
considered independently. According to Eqs.~(\ref{eq:sigmae}) and
(\ref{eq:sigmai}), $\sigma_{i}$ and $\sigma_{e}$ are both enhanced
when $a=a_{*}$, however variations in $\sigma_{e}$ may exceed those
of $\sigma_{i}$ by orders of magnitude which impacts the
experimental observables in a nontrivial way. Applying
this model to the experimental results of
Refs.~\cite{Pollack09,Machtey11,Zaccanti09}, we show that main
controversies in the second experimental approach can be resolved
when these processes are correctly included. We clarify the Efimov
resonance' positions in $^7$Li and $^{39}$K and support the emerging
evidence for significant finite range corrections to the first excited
Efimov energy level in both systems~\cite{Machtey11}.

In the following we assume that the number of dimers is always very
small compared to the number of trapped atoms and thus we neglect collisions
between them. This situation is correct for all the
experimental realizations considered here. We start with presenting
two possible scenarios ({\bf Sc}) that a dimer can undergo on
its way out of the trap:
\begin{description}
\item[Sc1:] The dimer collides elastically with $k$ atoms and leaves
the trap due to a high enough kinetic energy.
\item[Sc2:] After $k$ elastic collisions, the dimer undergoes one
inelastic collision, decays to a deeper bound state and leaves the trap without any
further scattering events because $\sigma_{e}$ and $\sigma_{i}$ are
enhanced only for the weakly bound universal dimers.
\end{description}
To evaluate the probabilities of each {\bf Sc}, we divide the total
length $l$ of the dimers' journey through the atomic cloud into $N$
segments of  infinitesimally small length $\delta l$. Then, most
generally, the probability to have an elastic/inelastic collision in
the length segment $\delta l$ is: $p_{e/i}=\sigma_{e/i} \overline{n}
\delta l$, where $\overline{n}$ is the mean density of
atoms~\footnote{For the sake of simplicity we consider column
density as the mean density ($\overline{n}$) multiplied by the mean
path-length of the dimer through the atom cloud ($l$) instead of the
mean column density $\overline{nl}$~\cite{Schuster01}. The
difference between the two is a constant of order $1$ and will have
negligible effect on the results presented here.}. Consequently, the
probability to have no events in the same segment is:
$p_{none}=1-p_{i}-p_{e}$.

According to Eq.~(\ref{eq:sigmai}), $\sigma_{i}$ depends on velocity
and thus changes at each elastic collision because the dimer
transfers a part of its initial kinetic energy to an atom. However,
it is instructive to start with the assumption that $\sigma_{i}$
remains constant independent of energy which makes the model clearer
and analytically more tractable. Moreover, this simplification is
reasonable in the limit of few collisional events and high initial
energies of the dimer. Experimentally it is realized in the region
of relatively small scattering lengths and moderate atom densities
which corresponds to some of the experimental conditions considered
later on.

\begin{figure}
{\centering \resizebox*{0.44\textwidth}{0.22\textheight}
{{\includegraphics{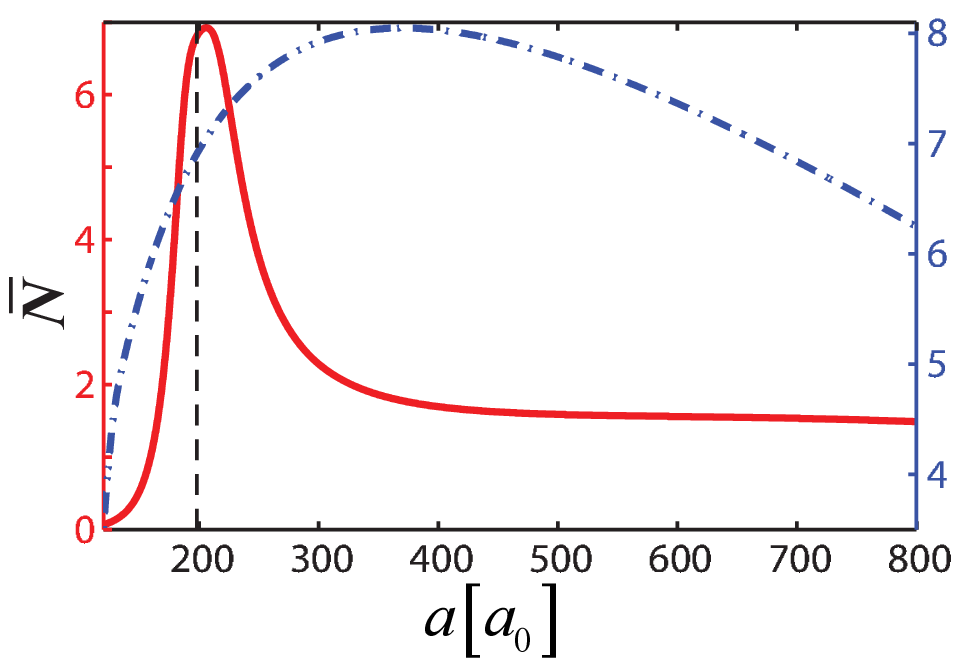}}}
\par}
\caption{\label{fig:Li} $\overline{N}(a)$ for the experimental
parameters of Ref.~\cite{Gross10} (solid red line) and
Ref.~\cite{Pollack09} (dashed-dotted line). Parameters of the Efimov
resonance are $a_{*}=196a_{0}$ (marked with the dashed vertical
line) and $\eta_{*}=0.1$.}
\end{figure}

{\it Energy independent (analytical) model.}

{\bf Sc1}: The probability for $k$ elastic events to occur can be
expressed by the choice of $k$ segments out of total N, ${N\choose
k}$, in which the elastic scatterings occur with the probability
$p_{e}^{k}$ and multiplying it by the probability $p_{none}^{N-k}$
that in the remaining $(N-k)$ segments no events happen. This
probability is given by:
\begin{equation}
p_{k,0}={N\choose k}p_{e}^{k}\;p_{none}^{N-k} \underset{\delta
l\rightarrow 0}\longrightarrow P\left(k;\sigma_{e}\overline{n} l
\right)\exp\left(-\sigma_{i}\overline{n} l \right).\label{eq:pse1s}
\end{equation}
where $P\left(k;x\right)$ is the Poisson distribution. Note that
although an inelastic event is not involved in Sc1, the probability
$p_{k,0}$ does depend on $\sigma_{i}$. This is because the
probability of no inelastic event has to be included as a factor.

{\bf Sc2}: The inelastic event, which happens after $k$ elastic
collisions, has to occur at any of the $N-k$ last segments with the
probability $p_{i}$. First, we fix the inelastic event at the $m-th$
place where $m\in\left[k+1,N\right]$. Then, the procedure is similar
to that of {\bf Sc1}. The probability for $k$ elastic collisions to
occur is expressed by the choice of $k$ segments out of total $m-1$,
${m-1\choose k}$, with the probability $p_{e}^{k}$ multiplied by the
probability $p_{none}^{m-1-k}$ that in the remaining $m-1-k$
segments no scattering events occur. Thus, the total probability for
{\bf Sc2} is just the sum over all possible locations of $m$:
\begin{eqnarray}
&&p_{k,1}=\sum_{m=k+1}^{N}{m-1\choose k}\;p_{none}^{m-1-k}\;p_{e}^{k}\;p_{i}\nonumber\\
&&\underset{\delta l\rightarrow 0}\longrightarrow \;
\int_{0}^{l}\frac{\left(\sigma_{e} \overline{n} x\right)^{k}}{k!}
\exp\left(-\left(\sigma_{e}+\sigma_{i}\right)\overline{n} x\right)\sigma_{i} \overline{n} \; dx.\label{eq:pse3s}
\end{eqnarray}

The mean number of scattering events in each {\bf Sc} is then
evaluated as the sum of weighted probabilities~\footnote{Note that the sum of all probabilities is exactly 1.}:
\begin{eqnarray}
\overline{N}_{e,0} &=&\sum_{k=1}^{\infty} k p_{k,0}=\sigma_{e}\overline{n} l \exp\left(-\sigma_{i}\overline{n} l\right),\label{eq:Nse1}\\
\overline{N}_{e,i} &=& \sum_{k=0}^{\infty}\left(k+1\right)p_{k,1}=
\left(1-\exp\left(-\sigma_{i}\overline{n} l\right)\right)\nonumber \\
&+&\frac{\sigma_{e}}{\sigma_{i}}
\left(1-\exp\left(-\sigma_{i}\overline{n} l
\right)\left(1+\sigma_{i}\overline{n} l \right)\right).
\label{eq:Nse2}
\end{eqnarray}
Finally, the total mean number of events is then: $\overline{N}
=\overline{N}_{e,0} +\overline{N}_{e,i}$. The central result of the
paper is already seen in Eqs.~(\ref{eq:Nse1},\ref{eq:Nse2}): both of
them include the interplay between elastic and inelastic cross
sections which strongly affects the experimental observables.

{\it Energy dependent model.} Now we revamp the model taking into
account the energy dependence of $\sigma_{i}$. When a dimer is created
via a three-body recombination, its energy is significantly larger
than the kinetic energy of atoms. The mean energy of the dimer is
reduced by a factor of $1/\alpha^2=5/9$ per elastic collision with a
nearly stationary atom~\cite{Fagnan09}. As before, we calculate the
probabilities and the mean number of events for each {\bf Sc}.

{\bf Sc1:} Let us consider $k$ elastic events. Then the first one
can occur only at any of the first $N-\left(k-1\right)$ segments
with the probability $p_{e}$. We fix its location at the segment
$j_{1}$ leaving the remaining $j_{1}-1$ segments to have no
scattering events with the probability
$\left(1-p_{i}\alpha^{0}-p_{e}\right)^{j_{1}-1}$. Next, we fix the
second scattering position $j_{2}$ which then can only occur between
segments $j_{1}+1$ and $N-\left(k-2\right)$ with the probability
$p_{e}$. In the remaining $\left(j_{2}-1\right)-j_{1}$ segments no
events happen with the probability
$\left(1-p_{i}\alpha^{1}-p_{e}\right)^{j_{2}-j_{1}-1}$ and so on
until the location $j_{k}$ is fixed. Finally, the total probability
of $k$ elastic events is evaluated by the sum of all these cases:
\begin{widetext}
\begin{equation}
p_{k,0}=p_{e}^{k}\sum_{j_{1}=1}^{N-k+1}
\;\sum_{j_{2}=j_{1}+1}^{N-k+2}
\cdots\underset{j_{k}=j_{k-1}+1}{\overset{N}{\sum}}
\prod_{m=1}^{k}\left(1-p_{i} \alpha^{m-1}-p_{e}\right)^{j_{m}-j_{m-1}-1}
\left(1-p_{i}
\alpha^{k}-p_{e}\right)^{N-k}, \text{ where $j_{0}=0$.} \label{eq:pse1f}
\end{equation}
\end{widetext}

{\bf Sc2:} Here the probability is constructed almost identically to
that of {\bf Sc1} but with one variation. Before counting the
elastic events we fix the segment where a single inelastic event
occurs at the $m-th$ position where $m\in\left[k+1,N\right]$. As a
result, elastic events can happen only in the first $m-1$ segments
and we count them as in {\bf Sc1}. Thus, the total probability for
{\bf Sc2} is just the sum over all possible locations of $m$:

\begin{widetext}
\begin{equation}
p_{k,1}=p_{e}^{k}p_{i}
\sum_{m=k+1}^{N}\;\sum_{j_{1}=1}^{m-k}\;\sum_{j_{2}=j_{1}+1}^{m-k+1}
\cdots\sum_{j_{k}=j_{k-1}+1}^{m-1}\prod_{s=1}^{k}
\left(1-p_{i}\alpha^{s-1}-p_{e}\right)^{j_{s}-j_{s-1}-1}
\left(1-p_{i}\alpha^{k}-p_{e}\right)^{m-k-1}\alpha^k, \label{eq:pse3f}
\end{equation}
\end{widetext}
where $j_{0}=0$. In the limit of $\delta l\rightarrow 0$, eqs.~(\ref{eq:pse1f},\ref{eq:pse3f}) take the form:

\begin{eqnarray}
p_{k,0}=\left(\frac{\sigma_{e}}{\sigma_{i}}\right)^{k}\sum_{j=0}^{k}
\frac{\exp\left(-\left(\sigma_{i}\alpha^{j}+\sigma_{e}\right)
\overline{n} l \right)}{\prod_{m=0,m\neq
j}^{k}\left(\alpha^{m}-\alpha^{j}\right)},
\qquad \\
p_{k,1}=\left(\frac{\sigma_{e}}{\sigma_{i}}\alpha\right)^{k}
\sum_{j=0}^k\frac{1-\exp\left(-\left(\sigma_{e}+\sigma_{i}\alpha^{j}\right) \overline{n} l \right)}{\left(\frac{\sigma_{e}}{\sigma_{i}}+\alpha^{j}\right)
\prod_{m=0\,,m\neq j}^k\left(\alpha^{m}-\alpha^{j}\right)}\;\;
\end{eqnarray}
which reduce to Eqs.~(\ref{eq:pse1s},\ref{eq:pse3s}) as
$\alpha\rightarrow 1$. The mean number of events $\overline{N}(a)$
can now be evaluated numerically based on the sum of weighted
probabilities (see Eqs.(\ref{eq:Nse1},\ref{eq:Nse2})). We now turn
to apply the model to the recent experimental results which use the second experimental strategy.

\emph{$^7Li$:} Ref.~\cite{Pollack09} reports $a_{*}=608a_{0}$, which
should be corrected to $a_{*}\sim 410a_{0}$ due to the shift in
position of the Feshbach resonance~\cite{Gross10,Gross11}. However,
recent measurement of Machtey $et.\:al.$ (MSGK)~\cite{Machtey11},
based on the same experimental technique reveals $a_{*}=196(4)a_{0}$
which is away from the Rice result by many times the experimental
uncertainty. There is, however, a major difference in the
experimental conditions used by the two groups: while MSGK work with
a low density thermal gas~\cite{Gross10,Machtey11}, the Rice
experiment is performed on a Bose-Einstein condensate (BEC).
According to Ref.~\cite{Pollack09}, the mean density of atoms in the
BEC is $\overline{n}\approx 5\times 10^{12}\;$cm$^{-3}$ and half of
the geometrical mean of the Thomas-Fermi radius is $l\approx
160\;\mu$m. In the MSGK case, for a typical temperature of
T=$1.4\;\mu$K, radial and longitudinal trap frequencies of
$\omega_{r}=2\pi\times 1.3\;$kHz and $\omega_{z}=2\pi\times 190\;$Hz
and $\sim 3.5\times 10^{4}$ atoms, the mean atom density is
$\overline{n}\approx 1\times 10^{12}\;$cm$^{-3}$. The geometrical
mean size of the atomic cloud is $l\approx 9.5\;\mu$m (one standard
deviation of the Gaussian distribution). Therefore the column
density ($\overline{n}l$) in the Rice experiment is larger by a
factor of $\sim 80$.

For the second strategy, all {\bf Sc}'s contribute to the
experimental observable and so we show $\overline{N}(a)$ in
Fig.~\ref{fig:Li} for both experiments. We used the energy dependent
model with $a_{*}=196a_{0}$. The solid red (dashed-dotted blue) line
corresponds to the MSGK (Rice) experimental conditions. In the MSGK
case, the maximum in $\overline{N}$ nearly perfectly coincides with
the position of $a_{*}$, which indicates that the experimental
observable indeed reveals the position of the Efimov resonance.
However, for the Rice experimental parameters, the maximum of
$\overline{N}$ is not at $a_{*}$ but rather appears at $\sim
375a_{0}$, in very good agreement with the reported
result~\cite{Pollack09}. A simple intuitive understanding of this
difference can be obtained when the limit of
$\sigma_{i}\overline{n}l\ll 1$ ($\gg 1$), roughly matching MSGK
(Rice) experimental conditions, is taken in
Eqs.~(\ref{eq:Nse1},\ref{eq:Nse2}). Then in the MSGK case,
$\overline{N}\approx \sigma_{e}\overline{n}l$, which corresponds to
the mean number of the elastic collisions and peaks at the maximum
of $\sigma_{e}$, i.e., at $a=a_{*}$. However, for the Rice case,
$\overline{N}\approx 1+\sigma_{e}/\sigma_{i}$, which is maximized at
a different position, namely when $s_{0}\ln(a/a_{*})+0.97=\pi/2$
(see Eqs.~(\ref{eq:sigmae},\ref{eq:sigmai}))~\footnote{This simple
condition is only obtained when $v \propto 1/a$ reflecting the
situation when the dimer is produced via three-body recombination
with the initial kinetic energy proportional to $E_{d}$.}. This
corresponds to $a/a_{*}\approx 1.8$ in agreement with the energy
dependent model and the Rice result.

Note that the model predicts a larger width of the $\overline{N}(a)$
maximum for the Rice experimental conditions, qualitatively well
identified in the experimental results~\cite{Pollack09,Machtey11}
and, thus, further strengthening our model. However, quantitative
comparison requires inclusion of additional conditions such as
reduction of the dimer energy below the trap depth after a certain
number of elastic collisions. Here we only associate the maximum in
the three-body recombination loss rate with the maximum in the mean
number of secondary collisions. In the region of large $a$ not all
of them, if any, lead to  atom loss, which cuts down the long tail
of $\overline{N}(a)$. This is, however, beyond the scope of the
present discussion.


\begin{figure}
{\centering \resizebox*{0.44\textwidth}{0.22\textheight}
{{\includegraphics{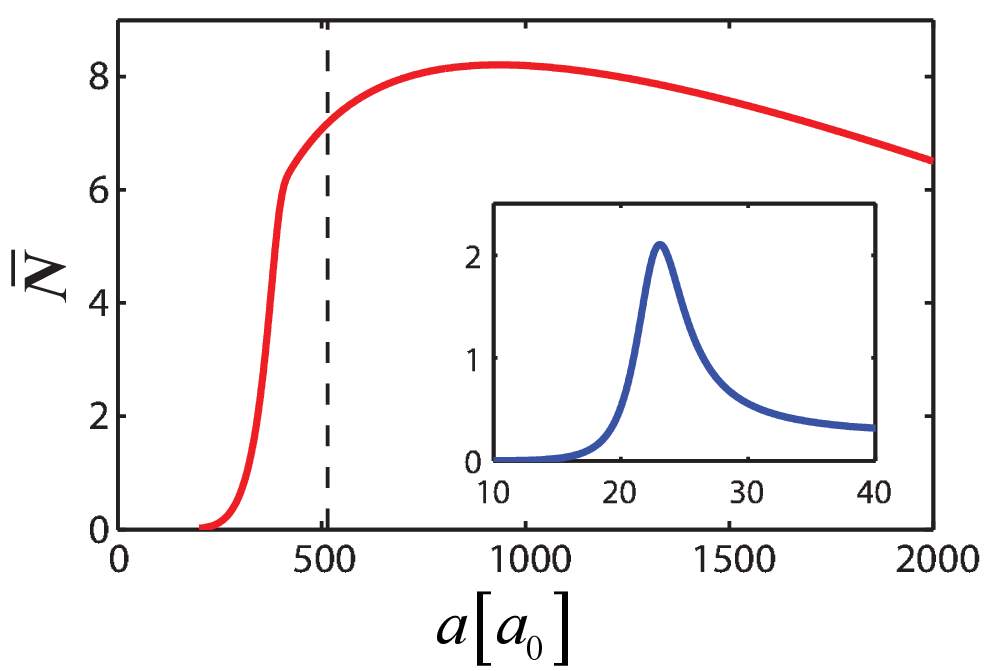}}}
\par}
\caption{\label{fig:K} $\overline{N}(a)$ for the experimental
parameters of Ref.~\cite{Zaccanti09}. Inset shows a peak at
low scattering length which also reasonably agrees with the experiment.
Parameters of the Efimov resonance are $a_{*}=515a_{0}$ (
marked with the dashed vertical line) and $\eta_{*}=0.1$.}
\end{figure}

\emph{$^{39}K$:} In Ref.~\cite{Zaccanti09} two Efimov resonances at
the atom-dimer threshold are reported at $a_{*1}=30.4a_{0}$ and
$a_{*2}=930a_{0}$. We shall concentrate our analysis on the second
resonance as the first one is deeply inside the nonuniversal region
where $a_{*1}<r_{0}$ with $r_{0}=64.5a_{0}$ being the van der Waals
length of $^{39}$K. The experiment is also performed in a BEC with
$\overline{n}\approx 6.5\times 10^{12}\;$cm$^{-3}$ and half of the
geometrical mean of the Thomas-Fermi radius is $l\approx 7\;\mu$m.
As before our model predicts a significant shift in the real
position of the resonance as compared to the reported one. To agree
with the experimental value of $\sim 930a_{0}$ we need to take
$a_{*}=515a_{0}$, which reflects the same numerical factor of $1.8$
as before. $\overline{N}(a)$ is shown in Fig.~\ref{fig:K}. In the
inset we represent the second peak located at $\sim 23a_{0}$ and
calculated with the slightly different experimental parameters in
accordance to Ref.~\cite{Zaccanti09} but with the same $a_{*}$.
Although it is close to the measured value it might be merely a
coincidence. Note also the difference in widths of the two features
which again qualitatively agrees with the experiment.

\begin{table}
\caption{\label{tab:sumtable} Comparison between newly extracted
values of the Efimov resonances at the atom-dimer threshold
$\left(a_{*}\right)$ and the values predicted by universal theory
$\left(a^{UT}_{*}\right)$. While $a^{M}_{*}$ denotes the originally
reported experimental results, $a_{*}$ shows the corrected values of
$a^{M}_{*}$ according to our model if applicable. $a^{*}_{0}$
indicates positions of recombination minima in the measured
three-body recombination spectra and is used to find $a^{UT}_{*}$
via a universal relation $a^{UT}_{*}\approx
1.1a^{*}_{0}\left(22.7\right)^{n-1/2}$ where $n$ can be either $0$
or $1$ for the purpose of this table~\cite{Braaten&Hammer06}.}
\begin{ruledtabular}
\begin{tabular}{ccccccc}
 &$a^{*}_{0} \left[a_{0}\right]$&$a^{UT}_{*} [a_{0}]$&$a^{M}_{*} [a_{0}]$&
 $a_{*} [a_{0}]$ &$a^{UT}_{*}/a_{*}$&$r_{0} [a_{0}]$\\
\hline $^7$Li~\cite{Machtey11}& $1260$\footnotemark[1] & $291$ & $196$ & $=a^{M}_{*}$ & $1.48$ & \multirow{2}{*}{$32.5$} \\
$^7$Li~\cite{Pollack09}\footnotemark[2]& $\sim1288$ & $\sim297$ & $\sim410$ & $\sim220$ & $\sim1.4$ & \\
\multirow{2}{*}{$^{39}$K~\cite{Zaccanti09}} & $224$ & $1174$ & \multirow{2}{*}{$930$} & \multirow{2}{*}{$515$} & $2.28$ & \multirow{2}{*}{$64.5$} \\
 & $5650$ & $1305$ &  &  & $2.53$ &   \\
$^{133}$Cs~\cite{Knoop09a}& $210$\footnotemark[3] & $1100$ & $397$ & $=a^{M}_{*}$ & $2.77$ & $101$ \\
\end{tabular}
\end{ruledtabular}
\footnotetext[1]{See also ref.~\cite{Gross10}.} \footnotetext[2]{The
values are rescaled according to correct Feshbach resonance
parameters, i.e. position and energy width~\cite{Gross10,Gross11}.}
\footnotetext[3]{See also refs.~\cite{Kraemer06,Berninger11}. Note
that $a^{*}_{0}$ is reasonably well related to an Efimov resonance
at $a<0$~\cite{Kraemer06} despite the fact that it is too close to
$r_{0}$ and thus might be subjected to the finite range
corrections.}
\end{table}

We can now compare the newly extracted positions of the Efimov
resonances at the atom-dimer threshold for both species with the
values predicted by universal theory (see Table~\ref{tab:sumtable}).
In both cases  $a_{*}$ is notably shifted to lower values. MSGK
associate this shift with the manifestation of finite range
corrections~\cite{Machtey11,Ji10}. For $^7$Li the shift is smaller
than for $^{39}$K. In both cases the Feshbach resonances are of
intermediate character between being closed and open channel
dominated, but $r_{0}$ of $^7$Li is about half that of $^{39}$K . It
is interesting to note that in Cs the shift is even larger despite
the open channel dominated character of the Feshbach resonance (see
Table~\ref{tab:sumtable})~\cite{Knoop09,Knoop09a,Kraemer06}.
However, the $r_{0}$ of Cs is the largest among the other species
and we can then conclude that the recent experimental results
support a scaling of the finite range corrections in $r_{0}$.
Finally, we note that experiments provide growing evidence for a
good agreement of the position of the lowest minimum in the
three-body recombination spectrum with universal theory despite the
fact that it is measured at even lower scattering lengths than
$a_{*}$~\cite{Kraemer06,Zaccanti09,Berninger11}, which still remains
a puzzling question.

This work was supported, in part, by the Israel Science Foundation.



\begin{thebibliography}{25}
\expandafter\ifx\csname natexlab\endcsname\relax\def\natexlab#1{#1}\fi
\expandafter\ifx\csname bibnamefont\endcsname\relax
  \def\bibnamefont#1{#1}\fi
\expandafter\ifx\csname bibfnamefont\endcsname\relax
  \def\bibfnamefont#1{#1}\fi
\expandafter\ifx\csname citenamefont\endcsname\relax
  \def\citenamefont#1{#1}\fi
\expandafter\ifx\csname url\endcsname\relax
  \def\url#1{\texttt{#1}}\fi
\expandafter\ifx\csname urlprefix\endcsname\relax\def\urlprefix{URL }\fi
\providecommand{\bibinfo}[2]{#2}
\providecommand{\eprint}[2][]{\url{#2}}

\bibitem[{\citenamefont{Braaten and Hammer}(2006)}]{Braaten&Hammer06}
\bibinfo{author}{\bibfnamefont{E.}~\bibnamefont{Braaten}} \bibnamefont{and}
  \bibinfo{author}{\bibfnamefont{H.-W.} \bibnamefont{Hammer}},
  \bibinfo{journal}{Phys. Rep.} \textbf{\bibinfo{volume}{428}},
  \bibinfo{pages}{259} (\bibinfo{year}{2006}).

\bibitem[{\citenamefont{Efimov}(1970)}]{Efimov70}
\bibinfo{author}{\bibfnamefont{V.}~\bibnamefont{Efimov}},
  \bibinfo{journal}{Phys. Lett. B} \textbf{\bibinfo{volume}{33}},
  \bibinfo{pages}{563} (\bibinfo{year}{1970}).

\bibitem[{\citenamefont{Ferlaino et~al.}(2011)\citenamefont{Ferlaino, Zenesini,
  Berninger, Huang, N\"{a}gerl, and Grimm}}]{Ferlaino11}
\bibinfo{author}{\bibfnamefont{F.}~\bibnamefont{Ferlaino}},
  \bibinfo{author}{\bibfnamefont{A.}~\bibnamefont{Zenesini}},
  \bibinfo{author}{\bibfnamefont{M.}~\bibnamefont{Berninger}},
  \bibinfo{author}{\bibfnamefont{B.}~\bibnamefont{Huang}},
  \bibinfo{author}{\bibfnamefont{H.-C.} \bibnamefont{N\"{a}gerl}},
  \bibnamefont{and} \bibinfo{author}{\bibfnamefont{R.}~\bibnamefont{Grimm}},
  \bibinfo{journal}{Few-Body Syst.} \textbf{\bibinfo{volume}{51}},
  \bibinfo{pages}{113} (\bibinfo{year}{2011}).

\bibitem[{\citenamefont{Kraemer et~al.}(2006)\citenamefont{Kraemer, Mark,
  Waldburger, Danzl, Chin, Engeser, Lange, Pilch, Jaakkola, N\"{a}gerl
  et~al.}}]{Kraemer06}
\bibinfo{author}{\bibfnamefont{T.}~\bibnamefont{Kraemer}},
  \bibinfo{author}{\bibfnamefont{M.}~\bibnamefont{Mark}},
  \bibinfo{author}{\bibfnamefont{P.}~\bibnamefont{Waldburger}},
  \bibinfo{author}{\bibfnamefont{J.~G.} \bibnamefont{Danzl}},
  \bibinfo{author}{\bibfnamefont{C.}~\bibnamefont{Chin}},
  \bibinfo{author}{\bibfnamefont{B.}~\bibnamefont{Engeser}},
  \bibinfo{author}{\bibfnamefont{A.~D.} \bibnamefont{Lange}},
  \bibinfo{author}{\bibfnamefont{K.}~\bibnamefont{Pilch}},
  \bibinfo{author}{\bibfnamefont{A.}~\bibnamefont{Jaakkola}},
  \bibinfo{author}{\bibfnamefont{H.-C.} \bibnamefont{N\"{a}gerl}},
  \bibnamefont{and} \bibinfo{author}{\bibfnamefont{R.}~\bibnamefont{Grimm}},
  \bibinfo{journal}{Nature}
  \textbf{\bibinfo{volume}{440}}, \bibinfo{pages}{315} (\bibinfo{year}{2006}).

\bibitem[{\citenamefont{Berninger et~al.}(2011)\citenamefont{Berninger,
  Zenesini, Huang, Harm, N{\"a}gerl, Ferlaino, Grimm, Julienne, and
  Hutson}}]{Berninger11}
\bibinfo{author}{\bibfnamefont{M.}~\bibnamefont{Berninger}},
  \bibinfo{author}{\bibfnamefont{A.}~\bibnamefont{Zenesini}},
  \bibinfo{author}{\bibfnamefont{B.}~\bibnamefont{Huang}},
  \bibinfo{author}{\bibfnamefont{W.}~\bibnamefont{Harm}},
  \bibinfo{author}{\bibfnamefont{H.-C.} \bibnamefont{N{\"a}gerl}},
  \bibinfo{author}{\bibfnamefont{F.}~\bibnamefont{Ferlaino}},
  \bibinfo{author}{\bibfnamefont{R.}~\bibnamefont{Grimm}},
  \bibinfo{author}{\bibfnamefont{P.~S.} \bibnamefont{Julienne}},
  \bibnamefont{and} \bibinfo{author}{\bibfnamefont{J.~M.}
  \bibnamefont{Hutson}}, \bibinfo{journal}{Phys.~Rev.~Lett.}
  \textbf{\bibinfo{volume}{107}}, \bibinfo{pages}{120401}
  (\bibinfo{year}{2011}).

\bibitem[{\citenamefont{Zaccanti et~al.}(2009)\citenamefont{Zaccanti, Deissler,
  D'Errico, Fattori, Jona-Lasinio, M\"{u}ller, Roati, Inguscio, and
  Modugno}}]{Zaccanti09}
\bibinfo{author}{\bibfnamefont{M.}~\bibnamefont{Zaccanti}},
  \bibinfo{author}{\bibfnamefont{B.}~\bibnamefont{Deissler}},
  \bibinfo{author}{\bibfnamefont{C.}~\bibnamefont{D'Errico}},
  \bibinfo{author}{\bibfnamefont{M.}~\bibnamefont{Fattori}},
  \bibinfo{author}{\bibfnamefont{M.}~\bibnamefont{Jona-Lasinio}},
  \bibinfo{author}{\bibfnamefont{S.}~\bibnamefont{M\"{u}ller}},
  \bibinfo{author}{\bibfnamefont{G.}~\bibnamefont{Roati}},
  \bibinfo{author}{\bibfnamefont{M.}~\bibnamefont{Inguscio}}, \bibnamefont{and}
  \bibinfo{author}{\bibfnamefont{G.}~\bibnamefont{Modugno}},
  \bibinfo{journal}{Nature Phys.} \textbf{\bibinfo{volume}{5}},
  \bibinfo{pages}{586} (\bibinfo{year}{2009}).

\bibitem[{\citenamefont{Gross et~al.}(2009)\citenamefont{Gross, Shotan,
  Kokkelmans, and Khaykovich}}]{Gross09}
\bibinfo{author}{\bibfnamefont{N.}~\bibnamefont{Gross}},
  \bibinfo{author}{\bibfnamefont{Z.}~\bibnamefont{Shotan}},
  \bibinfo{author}{\bibfnamefont{S.}~\bibnamefont{Kokkelmans}},
  \bibnamefont{and}
  \bibinfo{author}{\bibfnamefont{L.}~\bibnamefont{Khaykovich}},
  \bibinfo{journal}{Phys.~Rev.~Lett.} \textbf{\bibinfo{volume}{103}},
  \bibinfo{pages}{163202} (\bibinfo{year}{2009}).

\bibitem[{\citenamefont{Gross et~al.}(2010)\citenamefont{Gross, Shotan,
  Kokkelmans, and Khaykovich}}]{Gross10}
\bibinfo{author}{\bibfnamefont{N.}~\bibnamefont{Gross}},
  \bibinfo{author}{\bibfnamefont{Z.}~\bibnamefont{Shotan}},
  \bibinfo{author}{\bibfnamefont{S.}~\bibnamefont{Kokkelmans}},
  \bibnamefont{and}
  \bibinfo{author}{\bibfnamefont{L.}~\bibnamefont{Khaykovich}},
  \bibinfo{journal}{Phys.~Rev.~Lett.} \textbf{\bibinfo{volume}{105}},
  \bibinfo{pages}{103203} (\bibinfo{year}{2010}).

\bibitem[{\citenamefont{Pollack et~al.}(2009)\citenamefont{Pollack, Dries, and
  Hulet}}]{Pollack09}
\bibinfo{author}{\bibfnamefont{S.~E.} \bibnamefont{Pollack}},
  \bibinfo{author}{\bibfnamefont{D.}~\bibnamefont{Dries}}, \bibnamefont{and}
  \bibinfo{author}{\bibfnamefont{R.~G.} \bibnamefont{Hulet}},
  \bibinfo{journal}{Science} \textbf{\bibinfo{volume}{326}},
  \bibinfo{pages}{1683} (\bibinfo{year}{2009}).

\bibitem[{\citenamefont{Ottenstein et~al.}(2008)\citenamefont{Ottenstein,
  Lompe, Kohnen, Wenz, and Jochim}}]{Ottenstein08}
\bibinfo{author}{\bibfnamefont{T.~B.} \bibnamefont{Ottenstein}},
  \bibinfo{author}{\bibfnamefont{T.}~\bibnamefont{Lompe}},
  \bibinfo{author}{\bibfnamefont{M.}~\bibnamefont{Kohnen}},
  \bibinfo{author}{\bibfnamefont{A.~N.} \bibnamefont{Wenz}}, \bibnamefont{and}
  \bibinfo{author}{\bibfnamefont{S.}~\bibnamefont{Jochim}},
  \bibinfo{journal}{Phys.~Rev.~Lett.} \textbf{\bibinfo{volume}{101}},
  \bibinfo{pages}{203202} (\bibinfo{year}{2008}).

\bibitem[{\citenamefont{Huckans et~al.}(2009)\citenamefont{Huckans, Williams,
  Hazlett, Stites, and O'Hara}}]{Huckans09}
\bibinfo{author}{\bibfnamefont{J.~H.} \bibnamefont{Huckans}},
  \bibinfo{author}{\bibfnamefont{J.~R.} \bibnamefont{Williams}},
  \bibinfo{author}{\bibfnamefont{E.~L.} \bibnamefont{Hazlett}},
  \bibinfo{author}{\bibfnamefont{R.~W.} \bibnamefont{Stites}},
  \bibnamefont{and} \bibinfo{author}{\bibfnamefont{K.~M.}
  \bibnamefont{O'Hara}}, \bibinfo{journal}{Phys.~Rev.~Lett.}
  \textbf{\bibinfo{volume}{102}}, \bibinfo{pages}{165302}
  (\bibinfo{year}{2009}).

\bibitem[{\citenamefont{Williams et~al.}(2009)\citenamefont{Williams, Hazlett,
  Huckans, Stites, Zhang, and OHara}}]{Williams09}
\bibinfo{author}{\bibfnamefont{J.~R.} \bibnamefont{Williams}},
  \bibinfo{author}{\bibfnamefont{E.~L.} \bibnamefont{Hazlett}},
  \bibinfo{author}{\bibfnamefont{J.~H.} \bibnamefont{Huckans}},
  \bibinfo{author}{\bibfnamefont{R.~W.} \bibnamefont{Stites}},
  \bibinfo{author}{\bibfnamefont{Y.}~\bibnamefont{Zhang}}, \bibnamefont{and}
  \bibinfo{author}{\bibfnamefont{K.~M.} \bibnamefont{O'Hara}},
  \bibinfo{journal}{Phys.~Rev.~Lett.} \textbf{\bibinfo{volume}{103}},
  \bibinfo{pages}{130404} (\bibinfo{year}{2009}).

\bibitem[{\citenamefont{Lompe et~al.}(2010)\citenamefont{Lompe, Ottenstein,
  Serwane, Viering, Wenz, Z\"{u}rn, and Jochim}}]{Lompe10}
\bibinfo{author}{\bibfnamefont{T.}~\bibnamefont{Lompe}},
  \bibinfo{author}{\bibfnamefont{T.~B.} \bibnamefont{Ottenstein}},
  \bibinfo{author}{\bibfnamefont{F.}~\bibnamefont{Serwane}},
  \bibinfo{author}{\bibfnamefont{K.}~\bibnamefont{Viering}},
  \bibinfo{author}{\bibfnamefont{A.~N.} \bibnamefont{Wenz}},
  \bibinfo{author}{\bibfnamefont{G.}~\bibnamefont{Z\"{u}rn}}, \bibnamefont{and}
  \bibinfo{author}{\bibfnamefont{S.}~\bibnamefont{Jochim}},
  \bibinfo{journal}{Phys.~Rev.~Lett.} \textbf{\bibinfo{volume}{105}},
  \bibinfo{pages}{103201} (\bibinfo{year}{2010}).

\bibitem[{\citenamefont{Wild et~al.}()\citenamefont{Wild, Makotyn, Pino,
  Cornell, and Jin}}]{Wild11}
\bibinfo{author}{\bibfnamefont{R.~J.} \bibnamefont{Wild}},
  \bibinfo{author}{\bibfnamefont{P.}~\bibnamefont{Makotyn}},
  \bibinfo{author}{\bibfnamefont{P.~M.} \bibnamefont{Pino}},
  \bibinfo{author}{\bibfnamefont{E.~A.} \bibnamefont{Cornell}},
  \bibnamefont{and} \bibinfo{author}{\bibfnamefont{D.~S.} \bibnamefont{Jin}},
  \eprint{arXiv:1112.0362}.

\bibitem[{\citenamefont{Wang et~al.}()\citenamefont{Wang, D'Incao, Esry, and
  Greene}}]{Wang11}
\bibinfo{author}{\bibfnamefont{J.}~\bibnamefont{Wang}},
  \bibinfo{author}{\bibfnamefont{J.~P.} \bibnamefont{D'Incao}},
  \bibinfo{author}{\bibfnamefont{B.~D.} \bibnamefont{Esry}}, \bibnamefont{and}
  \bibinfo{author}{\bibfnamefont{C.~H.} \bibnamefont{Greene}},
  \eprint{arXiv:1201.1176}.

\bibitem[{\citenamefont{Chin}()}]{Chin11}
\bibinfo{author}{\bibfnamefont{C.}~\bibnamefont{Chin}},
  \eprint{arXiv:1111.1484}.

\bibitem[{\citenamefont{Knoop et~al.}(2009{\natexlab{a}})\citenamefont{Knoop,
  Ferlaino, Berninger, Mark, N\"{a}gerl, and Grimm}}]{Knoop09}
\bibinfo{author}{\bibfnamefont{S.}~\bibnamefont{Knoop}},
  \bibinfo{author}{\bibfnamefont{F.}~\bibnamefont{Ferlaino}},
  \bibinfo{author}{\bibfnamefont{M.}~\bibnamefont{Berninger}},
  \bibinfo{author}{\bibfnamefont{M.}~\bibnamefont{Mark}},
  \bibinfo{author}{\bibfnamefont{H.-C.} \bibnamefont{N\"{a}gerl}},
  \bibnamefont{and} \bibinfo{author}{\bibfnamefont{R.}~\bibnamefont{Grimm}},
  \bibinfo{journal}{Nature Phys.} \textbf{\bibinfo{volume}{5}},
  \bibinfo{pages}{227} (\bibinfo{year}{2009}{\natexlab{a}}).

\bibitem[{\citenamefont{Knoop et~al.}(2009{\natexlab{b}})\citenamefont{Knoop,
  Ferlaino, Berninger, Mark, N\"{a}gerl, and Grimm}}]{Knoop09a}
\bibinfo{author}{\bibfnamefont{S.}~\bibnamefont{Knoop}},
  \bibinfo{author}{\bibfnamefont{F.}~\bibnamefont{Ferlaino}},
  \bibinfo{author}{\bibfnamefont{M.}~\bibnamefont{Berninger}},
  \bibinfo{author}{\bibfnamefont{M.}~\bibnamefont{Mark}},
  \bibinfo{author}{\bibfnamefont{H.-C.} \bibnamefont{N\"{a}gerl}},
  \bibnamefont{and} \bibinfo{author}{\bibfnamefont{R.}~\bibnamefont{Grimm}},
  \bibinfo{journal}{J. Phys.: Conf. Ser.} \textbf{\bibinfo{volume}{194}},
  \bibinfo{pages}{012064} (\bibinfo{year}{2009}{\natexlab{b}}).

\bibitem[{\citenamefont{Nakajima et~al.}(2010)\citenamefont{Nakajima,
  Horikoshi, Mukaiyama, Naidon, and Ueda}}]{Nakajima10}
\bibinfo{author}{\bibfnamefont{S.}~\bibnamefont{Nakajima}},
  \bibinfo{author}{\bibfnamefont{M.}~\bibnamefont{Horikoshi}},
  \bibinfo{author}{\bibfnamefont{T.}~\bibnamefont{Mukaiyama}},
  \bibinfo{author}{\bibfnamefont{P.}~\bibnamefont{Naidon}}, \bibnamefont{and}
  \bibinfo{author}{\bibfnamefont{M.}~\bibnamefont{Ueda}},
  \bibinfo{journal}{Phys.~Rev.~Lett.} \textbf{\bibinfo{volume}{105}},
  \bibinfo{pages}{023201} (\bibinfo{year}{2010}).

\bibitem[{\citenamefont{Nakajima et~al.}(2011)\citenamefont{Nakajima,
  Horikoshi, Mukaiyama, Naidon, and Ueda}}]{Nakajima11}
\bibinfo{author}{\bibfnamefont{S.}~\bibnamefont{Nakajima}},
  \bibinfo{author}{\bibfnamefont{M.}~\bibnamefont{Horikoshi}},
  \bibinfo{author}{\bibfnamefont{T.}~\bibnamefont{Mukaiyama}},
  \bibinfo{author}{\bibfnamefont{P.}~\bibnamefont{Naidon}}, \bibnamefont{and}
  \bibinfo{author}{\bibfnamefont{M.}~\bibnamefont{Ueda}},
  \bibinfo{journal}{Phys.~Rev.~Lett.} \textbf{\bibinfo{volume}{106}},
  \bibinfo{pages}{143201} (\bibinfo{year}{2011}).

\bibitem[{\citenamefont{Braaten and Hammer}(2007)}]{Braaten&Hammer07}
\bibinfo{author}{\bibfnamefont{E.}~\bibnamefont{Braaten}} \bibnamefont{and}
  \bibinfo{author}{\bibfnamefont{H.-W.} \bibnamefont{Hammer}},
  \bibinfo{journal}{Ann. Phys.} \textbf{\bibinfo{volume}{322}},
  \bibinfo{pages}{120} (\bibinfo{year}{2007}).

\bibitem[{\citenamefont{Gross et~al.}(2011)\citenamefont{Gross,
Shotan, Machtey, Kokkelmans, and Khaykovich}}]{Gross11}
\bibinfo{author}{\bibfnamefont{N.}~\bibnamefont{Gross}},
  \bibinfo{author}{\bibfnamefont{Z.}~\bibnamefont{Shotan}},
  \bibinfo{author}{\bibfnamefont{O.}~\bibnamefont{Machtey}},
  \bibinfo{author}{\bibfnamefont{S.}~\bibnamefont{Kokkelmans}},
  \bibnamefont{and}
  \bibinfo{author}{\bibfnamefont{L.}~\bibnamefont{Khaykovich}},
  \bibinfo{journal}{C.~R.~Physique} \textbf{\bibinfo{volume}{12}},
  \bibinfo{pages}{4} (\bibinfo{year}{2011}).

\bibitem[{\citenamefont{Machtey et~al.}()\citenamefont{Machtey, Shotan, Gross,
  and Khaykovich}}]{Machtey11}
\bibinfo{author}{\bibfnamefont{O.}~\bibnamefont{Machtey}},
  \bibinfo{author}{\bibfnamefont{Z.}~\bibnamefont{Shotan}},
  \bibinfo{author}{\bibfnamefont{N.}~\bibnamefont{Gross}}, \bibnamefont{and}
  \bibinfo{author}{\bibfnamefont{L.}~\bibnamefont{Khaykovich}},
  \eprint{arXiv:1201.2396}.

\bibitem[{\citenamefont{Fagnan et~al.}(2009)\citenamefont{Fagnan, Wang, Zhu,
  Djuricanin, Klappauf, Booth, and Madison}}]{Fagnan09}
\bibinfo{author}{\bibfnamefont{D.~E.} \bibnamefont{Fagnan}},
  \bibinfo{author}{\bibfnamefont{J.}~\bibnamefont{Wang}},
  \bibinfo{author}{\bibfnamefont{C.}~\bibnamefont{Zhu}},
  \bibinfo{author}{\bibfnamefont{P.}~\bibnamefont{Djuricanin}},
  \bibinfo{author}{\bibfnamefont{B.~G.} \bibnamefont{Klappauf}},
  \bibinfo{author}{\bibfnamefont{J.~L.} \bibnamefont{Booth}}, \bibnamefont{and}
  \bibinfo{author}{\bibfnamefont{K.~W.} \bibnamefont{Madison}},
  \bibinfo{journal}{Phys. Rev. A} \textbf{\bibinfo{volume}{80}},
  \bibinfo{pages}{022712} (\bibinfo{year}{2009}).

\bibitem[{\citenamefont{Ji et~al.}(2010)\citenamefont{Ji, Phillips, and
  Platter}}]{Ji10}
\bibinfo{author}{\bibfnamefont{C.}~\bibnamefont{Ji}},
  \bibinfo{author}{\bibfnamefont{D.}~\bibnamefont{Phillips}}, \bibnamefont{and}
  \bibinfo{author}{\bibfnamefont{L.}~\bibnamefont{Platter}},
  \bibinfo{journal}{Europhys.~Lett.} \textbf{\bibinfo{volume}{92}},
  \bibinfo{pages}{13003} (\bibinfo{year}{2010}).

\bibitem[{\citenamefont{Schuster et~al.}(2001)\citenamefont{Schuster, Marte,
  Amtage, Sang, Rempe, and Beijerinck}}]{Schuster01}
\bibinfo{author}{\bibfnamefont{J.}~\bibnamefont{Schuster}},
  \bibinfo{author}{\bibfnamefont{A.}~\bibnamefont{Marte}},
  \bibinfo{author}{\bibfnamefont{S.}~\bibnamefont{Amtage}},
  \bibinfo{author}{\bibfnamefont{B.}~\bibnamefont{Sang}},
  \bibinfo{author}{\bibfnamefont{G.}~\bibnamefont{Rempe}}, \bibnamefont{and}
  \bibinfo{author}{\bibfnamefont{H.~C.~W.} \bibnamefont{Beijerinck}},
  \bibinfo{journal}{Phys. Rev. Lett.} \textbf{\bibinfo{volume}{87}},
  \bibinfo{pages}{170404} (\bibinfo{year}{2001}).

\end{thebibliography}

\end{document}